\documentclass{PoS}

\title{Gamma-ray emitting radio galaxies at hard X-rays: Seyfert core or jet emission?}

\ShortTitle{Gamma-ray bright radio galaxies}

\author{\speaker{Volker Beckmann}, Sandra de Jong, Fabio Mattana, Dounia Saez, \& Simona Soldi\\ 
        Fran\c{c}ois Arago Centre, APC, Universit\'e Paris Diderot, CNRS/IN2P3, CEA/Irfu, Observatoire de Paris, Sorbonne Paris Cit\'e, 13 rue Watt, 75013 Paris, France\\
        E-mail: \email{beckmann@apc.univ-paris7.fr}}

\abstract{A number of radio galaxies has been detected by Fermi/LAT in the gamma-ray domain. In some cases, like Cen~A and M~87, these objects have been seen even in the TeV range by Cherenkov telescopes. Whereas the gamma-ray emission is likely to be connected with the non-thermal jet emission, dominating also the radio band, the situation is less clear at hard X-rays. While the smoothly curved continuum emission and the overall spectral energy distribution indicate a non-thermal emission, other features such as the iron line emission and the low variability appear to be rather of Seyfert type, i.e. created in the accretion disk and corona around the central black hole. 

We investigate several prominent cases using combined X-ray and gamma-ray data in order to constrain the possible contributions of the jet and the accretion disk to the overall spectral energy distribution in radio galaxies. 
Among the three sources we study, three different origins of the hard X-ray flux can be identified. The emission can be purely non-thermal and caused by the jet, as in the case of M~87, or thermal inverse Compton emission from the Seyfert type core (Cen~A), or appears to be a superposition of non-thermal and thermal inverse Compton emission, as we observe in 3C~111.
Gamma-ray bright radio galaxies host all kinds of AGN cores, Seyfert 1  and 2, BL Lac objects, and also LINER.   
}

\FullConference{"An INTEGRAL view of the high-energy sky (the first 10 years)"
9th INTEGRAL Workshop and celebration of the 10th anniversary of the launch,\\
		October 15-19, 2012\\
		Bibliotheque Nationale de France, Paris, France}

\begin{document}

\section{Introduction}

Among the radio loud objects, radio galaxies present the cases where the central engine is hidden by circumnuclear mattter, but the object produces bright radio jets, sometimes with prominent extended lobes \cite{AGNbook}. In order to be detectable as a radio galaxy, the jet angle with respect to the line of sight has to be large, i.e. $\gg 10^\circ$ because for smaller angles the object would appear as a blazar in which we assume to look right down the jet. 
The main emission component we see in the extended images of radio galaxies is produced by synchrotron emission of the charged particles interacting with the magnetic field of the jet. The inverse Compton component of radio galaxies is expected to appear rather as that of Seyfert galaxies. The emission should not be efficiently beamed toward the observer, as the jet is not pointing towards us, resulting in a component decreasing rapidly in the X-ray domain, with a high-energy cut-off in the range 50--200 keV. 

One exception from this rule has been known since the observations of the {\it Compton Gamma-Ray Observatory} ({\it CGRO}) in the 1990s: Cen~A was detected by {\it CGRO}/COMPTEL and EGRET up to $\sim 1 \rm \, GeV$. With {\it Fermi}/LAT the number has increased and there are now more than a dozen gamma-ray bright radio galaxies known. Four radio galaxies, all of them of type FR-I, are detectable up to the TeV range (3C~66B, NGC~1275, M~87, Cen~A). This raises the question, how the high photon energies $E \gg 100 \rm \, keV$ are achieved in these cases, and what makes a radio galaxy gamma-ray bright \cite{Rieger12}. For example, the apparently brightest extragalactic radio source, the \mbox{FR-II} Cygnus~A is not a gamma-ray emitter, although it displays a hard X-ray flux of $f_{20 - 100 \rm \, keV} = 8 \times 10^{-11} \rm \, erg \, cm^{-2} \, s^{-1}$ \cite{AGN2nd}. 
Gamma-ray loud radio galaxies are of further interest as they might be responsible for a large fraction of the cosmic background above 1~MeV \cite{Inoue11}.

In this work we use X-ray data and the overall spectral energy distribution (SED) to investigate the characteristics of 3 gamma-ray bright radio galaxies: Cen~A, 3C~111, and M~87. In the X-rays we want to clarify whether the observed spectrum has thermal or non-thermal characteristics, i.e. whether it is produced by inverse Compton scattering close to the central engine in a electron plasma cloud (e.g. a hot corona on top of the accretion disk) or in a jet.

\section{Centaurus A}

Centaurus~A is a FR-I radio galaxy that has been detected all the way from the radio domain up to the TeV region. Its brightness and small distance ($D = 3.8 \rm \, Mpc$) makes it an ideal target to clarify in which energy bands the emission is thermal and where it appears to be dominated by non-thermal processes.
The double hump structure of the SED with an apparently continuous inverse Compton emission from the X-rays to the TeV domain suggests that this emission is of non-thermal origin. Analysis of {\it CGRO} data shows that the emission from 50 keV up to 1 MeV is well represented by a double broken power-law model \cite{Steinle98}, which could be a simplification of the smoothly curved continuous inverse Compton emission of a blazar and therefore caused by the jet. 
One has to keep in mind though, that in the range $0.9 - 3 \rm \, MeV$, the {\it CGRO} data provided only upper limits on the source flux. A transition from a thermal inverse Compton (IC) dominated process to a non-thermal component could therefore occur in this energy range and would have passed unnoticed. 

In order to clarify the situation we performed an in-depth analysis of all the available \mbox{\it INTEGRAL} data \cite{CenA}. The analysis of the 3--1000 keV data showed that the overall spectrum can be fit by a physical model of thermal Comptonisation (compPS) giving a plasma temperature of $kT_{e} =$ 206$\pm 62 \rm \, keV$ within the optically thin corona with Compton parameter $y = 0.42 {+0.09 \atop -0.06}$. A reflection component with a reflection strength $R = 0.12 {+0.09 \atop -0.10}$ has been measured. Therefore, the data are consistent with $R = 0$ on the $1.9\sigma$ level, and a strength of $R > 0.3$ can be excluded on a $3\sigma$ level. 
 Extending the cut-off power-law or the Comptonisation model to the gamma-ray range shows that they cannot account for the high-energy emission. On the other hand, a broken or curved power-law model can also represent the data, therefore a non-thermal origin of the X-ray to GeV emission could not be ruled out.

\begin{figure}
\begin{center}
\includegraphics[angle=0,width=.9\textwidth]{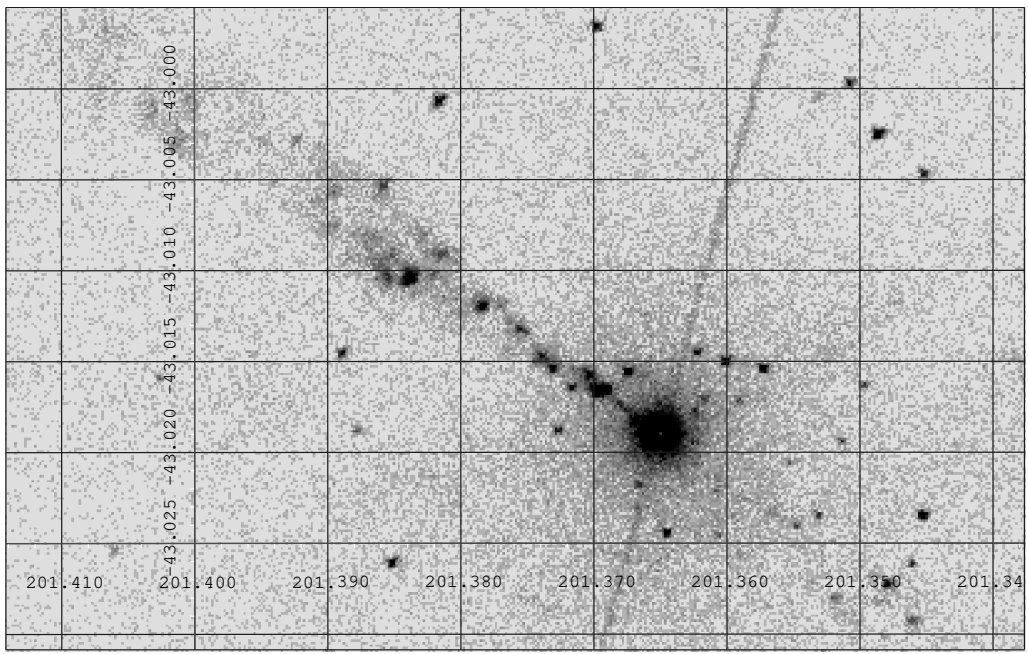}
\caption{{\it Chandra}/ACIS-I image of Centaurus~A based on a 98 ksec exposure taken in May 2007. The core, which is piled-up in this image, shows a 100 times higher flux level than the combined extended and jet emission (including the hot spots). For comparison: in M~87, which is another gamma-ray detected radio galaxy, the jet emits about 4 times as much as the core in the 2--10 keV band \cite{M87}.}
\label{fig:CenAimage}
\end{center}
\end{figure}

We then studied the extended, jet, and core emission in the 2--10 keV band as observed and resolved by {\it Chandra} (Fig.~\ref{fig:CenAimage}). 
Extrapolating the phenomenological models we fit to the core and jet spectra in the {\it Chandra} band to the 20--100 keV energy range covered by {\it INTEGRAL} IBIS/ISGRI, we find that the expected flux for the core emission of Cen~A is $f_{20 - 100 \rm \, keV} \simeq 6 \times 10^{-10} \rm \, erg \, cm^{-2} \, s^{-1}$, while the expected flux of the jet is about two orders of magnitude lower with $f_{20 - 100 \rm \, keV} \simeq 7\times 10^{-12} \rm \, erg \, cm^{-2} \, s^{-1}$. It turns out that the extrapolated core spectrum gives indeed the flux we measure in IBIS/ISGRI \cite{CenA}. Thus, we can conclude that the emission in the hard X-ray band is still dominated by the Seyfert-type core and can be described as thermal inverse Compton process. 

The transition to the non-thermal, jet dominated spectral energy distribution has then to take place somewhere between a few 100~keV and $\sim 10 \rm \, MeV$. This underlines the need for a dedicated satellite mission in this up to now poorly explored energy range of $1 - 100 \rm \, MeV$ \cite{Lebrun10}.

\section{3C 111}

3C~111 is a FR-II radio galaxy at redshift $z = 0.049$ ($D = 205 Mpc$), which displays a broad line region in the optical and iron K$\alpha$ emission in the X-rays, similar to Seyfert galaxies. The radio jet has an angle of $18^\circ$ to the line of sight. This object has been first detected by {\it Fermi}/LAT in the gamma-rays, with a photon index of $\Gamma = 2.4 \pm 0.2$ in the range $0.1 - 200 \rm \, GeV$ and a flux of $f \simeq 10^{-8} \rm \, ph \, cm^{-2} \, s^{-1}$ \cite{3C111}.
This low flux level allows only for a low significant detection of $\sim 3 \sigma$. In both, the soft and the hard X-rays, the source is fainter than Cen~A, with 
 $f_{20 - 200 \rm \, keV} \simeq 10^{-10} \rm \, erg \, cm^{-2} \, s^{-1}$. 

In order to determine whether the (hard) X-ray data are dominated by non-thermal or thermal emission, we analysed combined {\it Suzaku} XIS and PIN, {\it Swift}/BAT, and {\it INTEGRAL} IBIS/ISGRI data. The overall spectrum in the $0.4 - 200 \rm \, keV$ band shows an absorbed exponential cut-off power-law with
reflection from neutral material with a photon index $\Gamma = 1.68 \pm 0.03$, a high-energy cut-off at $E_{\rm cut} = 230 {+140 \atop -70} \rm \, keV$, a reflection component with $R = 0.7 \pm 0.3$ and an iron K$\alpha$ line with an equivalent width of $EW = 85 \pm 11 \rm \, eV$ \cite{3C111}. 
Thus, in the case of 3C~111 we see indication for both, thermal and non-thermal inverse Compton emission in the X-rays. The measured high-energy cut-off can occur in a thermal IC emission, or can be the signature of a smoothly curved hard X-ray spectrum. The iron line and the Compton reflection hump are clear indications for a thermal IC process, but at the same time the iron line is rather weak compared to the continuum emission and the overall spectral energy distribution can be fit by a one-zone synchrotron self-Compton model as typical for blazars. 
This seems to suggest that although there is compelling evidence for Seyfert-like emission from 3C~111, the X-rays are most likely also influenced by non-thermal jet emission. This is further supported by the fact that the equivalent width of the iron line and the continuum emission are variable, and we observe a highly-significant anti-correlation between the equivalent width and the continuum flux. It seems, that the variations of the underlying continuum by a factor of $\sim 4$ are driven by the variable non-thermal emission, while the Seyfert core remains persistent. 

Thus, for 3C~111 we conclude that the X-rays show contribution of both, non-thermal jet emission and thermal inverse Compton emission similar to that in Seyfert galaxies. Also here, like in the other radio galaxies, we assume that the gamma-ray emission can be entirely attributed to jet emission, because the thermal IC component with its exponential cut-off in the hard X-rays will not contribute significantly beyond $\simeq 500 \rm \, keV$.

\section{M 87}

The third radio galaxy we studied in this work is the FR-I M~87 at a distance of $D = 16 \rm \, Mpc$. In this case, similar as for Cen~A, the different components can be resolved in the optical and X-ray regime. The jet of M~87 has an angle to the line of sight of $30^\circ$, and several knots can be distinguished in the X-rays based on Chandra data \cite{Wilson02}. The source is detectable up to the TeV range, but so far has escaped detection in the hard X-rays above 10 keV, with a $3\sigma$ upper limit of $f < 3 \times 10^{-12} \rm \, erg \, cm^{-2} \, s^{-1}$ \cite{M87}. Different than in the case of Cen~A and 3C~111, this source shows a dominance of the jet over the core emission in the 2--10 keV band: the jet emission is on average about a factor of 4 brighter than the core. This ratio depends on the time of observation, as the jet emission is highly variable. In addition, there is no iron K$\alpha$ emission detectable from the core. Studying the hard X-ray light curve as derived from 1.7 Msec effective on-source observation with {\it INTEGRAL} IBIS/ISGRI, there is no detection achievable in the hard X-rays at any given time. 
The overall spectral energy distribution can be modeled, like in the case of 3C~111, by a one-zone synchrotron self-Compton model with parameters typical for high-frequency peaked BL Lac object. 

Further investigations on the nature of the SED and the possibility to detect M~87 at hard X-rays are performed and will be presented soon (de Jong et al. 2013, in preparation). But the picture which emerges for M~87 is that of a radio galaxy with no sign of Seyfert-type activity in the core detectable in X-rays, with this energy range entirely dominated by non-thermal emission. Considering that the core has been classified as that of a LINER in the optical domain \cite{M87optical}, the lack of strong thermal inverse Compton emission in the X-rays can be expected because of the low power output of the central engine, and LINER are also known for not displaying an iron K$\alpha$ line \cite{Younes11}. 
In summary, it seems that the X-ray emission of M~87 is dominated by non-thermal inverse Compton emission arising from the jet.

\section{Fermi/LAT detected radio galaxies in the X-rays}

Although we have investigated only three particular cases, already in this small sample we find three different types of X-ray emission in gamma-ray bright radio galaxies. The hard X-rays are either dominated by non-thermal emission from the jet (M~87; FR-I), by thermal inverse Compton emission arising from the Seyfert type core (Cen~A; FR-I), or a composite of core and jet emission is visible in this band (3C~111; FR-II). Summarizing some of the parameters of interest in Table~1, it might be that the core dominance in the X-rays in the case of Cen~A can be attributed to the large jet angle with respect to the line of sight. The fact that the thermal inverse Compton component is not visible in the case of M~87 could be caused by the relatively weak AGN core. 

\begin{table}
\centering
\begin{tabular}{c c c c c c c}
 & FR type & jet angle & core type & $f_{> 100 \rm \, MeV}$ & $f_{20-60 \rm \, keV}$ & X-rays\\
 &         &           &           & $[\rm ph \, cm^{-2} \, s^{-1}]$ & $[\rm erg \, cm^{-2} \, s^{-1}]$ & \\
\hline
M 87  & FR-I & $30^\circ$ & LINER & $2.4 \times 10^{-8}$ & $< 3 \times 10^{-12}$ & jet\\
Cen A & FR-I & $50^\circ - 80^\circ$ & Seyfert 1.5 & $2.1 \times 10^{-7}$ & $6 \times 10^{-10}$ & core\\
3C 111& FR-II& $18^\circ$ & Seyfert 1 &  $4 \times 10^{-8}$ & $8.5 \times 10^{-11}$ & core+jet
\label{comparison}
\end{tabular}
\caption{Selected parameters of the three gamma-ray bright radio galaxies studied here.}
\end{table}

Three objects are not sufficient in order to draw any firm conclusions here. In a next step we will investigate the hard X-ray spectra of other gamma-ray detected radio galaxies, in order to see whether a trend emerges in which certain physical parameters of the objects determine the gamma-ray brightness. 
Already a glance at the optical core classification of these galaxies seems to indicate that the type of AGN at the center does not give a hint whether or not a radio galaxy can become gamma-ray bright. Radio quiet AGN do not seem to be significant gamma-ray emitters \cite{Ackermann12}. For example, among 491 {\it Swift}/BAT detected Seyfert galaxies, only NGC~4945 and NGC~1068 were detected by {\it Fermi}/BAT, and in these cases the star burst component in the host galaxy is likely to be the origin of the high-energy emission \cite{Teng11}.
Among the {\it Fermi}/LAT detected radio galaxies, there are cores with BL~Lac spectra (IC~310), of type Seyfert~2 (NGC~1275, NGC~6251), Seyfert~1.5 (3C~380.0, Cen~A), Seyfert~1.2 (3C~207.0), Seyfert~1 (3C~111), several LINER (\mbox{Fornax~A}, OH~-342, M~87), and even an apparently optically non-active elliptical galaxy (Cen~B). Gamma-ray bright radio galaxies are also observed in a large redshift range, from Cen~A at a distance of $D = 3.8 \rm \, Mpc$ out to $2,400 \rm \, Mpc$ ($z = 0.68$) in the case of 3C~207.0, and thus we observe objects which are very different in terms of luminosity.\\

{\it Acknowledgements}: This work has been supported by the LabEx UnivEarthS\footnote{http://www.univearths.fr/en} project ``Impact of black holes on their environment''. We thank the anonymous referee for the constructive comments which helped to improve this paper.

\end{document}